\documentclass[12pt]{article}
\usepackage{latexsym}

\def\bs{\begin{subequations}}
\def\es{\end{subequations}}

\catcode`\@=11

\newtoks\@stequation
\def\subequations{\refstepcounter{equation}
  \edef\@savedequation{\the\c@equation}%
  \@stequation=\expandafter{\theequation}
  \edef\@savedtheequation{\the\@stequation}
  \edef\oldtheequation{\theequation}%
  \setcounter{equation}{0}%
  \def\theequation{\oldtheequation\alph{equation}}}

\def\endsubequations{\setcounter{equation}{\@savedequation}%
  \@stequation=\expandafter{\@savedtheequation}%
  \edef\theequation{\the\@stequation}\global\@ignoretrue}

\makeatletter
        \renewcommand{\theequation}{\thesection.\arabic{equation}}%
        \@addtoreset{equation}{section}%
\makeatother

\renewcommand{\thefootnote}{\fnsymbol{footnote}}

\begin{document}

\begin{titlepage}

 June 12, 2012

\begin{center}        \hfill   \\
            \hfill     \\
                                \hfill   \\

\vskip .25in

{\large \bf A Conjecture about Conserved Symmetric Tensors \\}

\vskip 0.3in

Charles Schwartz\footnote{E-mail: schwartz@physics.berkeley.edu}

\vskip 0.15in

{\em Department of Physics,
     University of California\\
     Berkeley, California 94720}
        
\end{center}

\vskip .3in

\vfill

\begin{abstract}
We consider T(x), a tensor of arbitrary rank  that is symmetric in all of its 
indices and 
conserved in the sense that the divergence on any one index vanishes. 
Our conjecture is that all integral moments of this tensor will 
vanish if the number of coordinates in that integral moment is less than the 
rank of the tensor. This result is proved 
explicitly for a number of particular cases, assuming adequate
dimensionality of the Euclidean space of coordinates (x); but 
a general proof is lacking. Along the way, we find some neat results for 
certain large matrices generated by permutations.

\end{abstract}

\vfill

\end{titlepage}

\renewcommand{\thefootnote}{\arabic{footnote}}
\setcounter{footnote}{0}
\renewcommand{\thepage}{\arabic{page}}
\setcounter{page}{1}

\section{Introduction}

In an n-dimensional real Euclidean space we consider symmetric 
tensors of any rank that are ``conserved'' as follows,
\begin{equation}
\sum_{i=1,n}\frac{\partial}{\partial x_{i}} \;T_{i w} (x) = 
0,\label{a1}
\end{equation}
where $w$ is a string of indices (a ``word'') of arbitrary length. If 
$w = abc$, then $iw = iabc$, etc.; and we use the notation $[w]$ to 
denote the length of the word $w$.

Our interest is in integrals of the form
\begin{equation}
(w_{L};w_{R}) \equiv \int d^{n}x\; x^{w_{L}}\;T_{w_{R}}(x),\label{a2}
\end{equation}
which we call ``integral moments'' of the tensor $T$, and $x^{w}$ is a product of coordinates identified by the letters 
in the word $w$. We denote the null word by $0$, so that $x^{0}=1$. 
We assume that the tensors are functions well confined in space,
so that 
all such integrals of interest converge and we can do partial 
integration ignoring surface terms.

Our conjecture is that all integrals of the type (\ref{a2}) will vanish 
so long as $[w_{L}] < [w_{R}]$; and the identities (\ref{a1}) are necessary for 
this result to be true. 

Usually one speaks about tensors in relation to some group of 
transformations; but here that plays no role. This is just about 
algebraic manipulation of the indices.

This topic arose from a recent study of the General Theory of 
Relativity \cite{CS}, where we were looking at the asymptotic form of 
the potential produced by a source represented by such a symmetric 
tensor. The main conclusion was that there is no  
long range potential $\sim 1/r$.

\section{The system of equations}

The general identity we start with is this,
\begin{equation}
<w_{1}|w_{2}> \equiv -\sum_{i} \int d^{n}x\; 
x^{w_{1}}\;\frac{\partial}{\partial x_{i}}\;
T_{i w_{2}}(x) =  
\sum_{i} (\partial_{i}w_{1};iw_{2}) = 0, \label{a3}
\end{equation}
where $\partial_{i}w$ means removing any occurence of the letter $i$ 
in the word $w$. Since we are interested in symmetric tensors, the order 
of letters in any word is unimportant.

We can organize this host of equations by looking at the combined 
word $W = w_{1}w_{2} = w_{L}w_{R}$; and we see that the equations (\ref{a3}) 
separate into distinct subsets for each combined word.

\section{Examples}

We proceed from the simplest examples of (\ref{a3}) to more complicated ones. In 
what follows, I use the letters $a,b,c,\ldots$ to denote distinct 
values of the index $i = 1,2,\ldots,n$. It will be advantageous to 
separate subsets of letters according to Young Tableaux, such as 
$a^{3}b^{2}cd$, for example. The word $w$, as used below may be  
arbitrary.

\begin{eqnarray}
[w_{L}]=0; W=aw: \\ 
 <a|w> = (0;aw) = 0.
\end{eqnarray}
This is the simplest case, which says, when written out,
\begin{equation}
 -\int d^{n}x\;x_{a}\sum_{i=1,n}\;\frac{\partial}{\partial x_{i}} T_{iw}(x) = \int 
d^{n}x\;T_{aw}(x) = 0.\label{a4}
\end{equation}

We proceed to $[w_{L}]=1$.
\begin{eqnarray}
[w_{L}]=1; W = a^{2}bw: \label{c}\\ 
 <aa|bw> = 2(a;abw) = 0, \\ 
 <ab|aw> = (a;abw) + (b;aaw) = 0. 
\end{eqnarray}
So all these are zero.

\begin{eqnarray}
[w_{L}]=1; W = abcw:\label{c2} \\ 
 <ab|cw> = (a;bcw) + (b;acw) = 0, \\ 
 <bc|aw> = (b;acw) + (c;abw) = 0, \\ 
 <ca|bw> = (c;abw) + (a;bcw) = 0.
\end{eqnarray}
The solution of these three simultaneous equations in three unknows 
is that all three are zero. We had to introduce the label $c$ 
explicitly to get this result; which means that the tensors involved 
here are at least of second rank and also that the dimensionality of 
the space must be $n\ge 3$.

Now we go to $[w_{L}]=2$; and here we need $[W]\ge 5$. The case $W = 
a^{5}w$ is trivial and $W=a^{4}bw$ is similar to what is above in 
(\ref{c}).
\begin{eqnarray}
[w_{L}]=2; W = a^{3}b^{2}w \\
<aaa|bbw> = 3(aa;abbw)=0, \\ 
<aab|abw> = 2(ab;aabw) + (aa;abbw)=0, \\
<abb|aaw> =  (bb;aaaw) + 2(ab|aabw)=0.
\end{eqnarray}
The solution is that all three unknowns are zero.

\begin{eqnarray}
[w_{L}]=2; W = a^{3}bcw \\
<aaa|bcw> = 3(aa;abcw)=0, \\ 
<aab|acw> = 2(ab;aacw) + (aa;abcw)=0, \\
<aac|abw> = 2(ac;aabw) + (aa;abcw)=0, \\
<abc|aaw> = (ab;aacw) + (ac;aabw) + (bc;aaaw)=0. \\
\end{eqnarray}
The solution is that all four unknowns are zero.

\begin{eqnarray}
[w_{L}]=2; W= a^{2}b^{2}cw: \\ 
<aab|bcw> = 2(ab;abcw) + (aa;bbcw)  = 0, \\ 
<abb|acw> = 2(ab;abcw) + (bb;aacw) = 0, \\ 
<bbc|aaw> = 2(bc;aabw) + (bb;aacw) = 0, \\
<abc|abw> =  (ab;abcw) + (ac;abbw) + (bc;aabw) = 0,\\ 
<aac|bbw> = 2(ac;abbw) + (aa;bbcw) = 0.  
\end{eqnarray}
These 5 equations in 5 unknowns have the solution that all  
unknowns are zero.

\begin{eqnarray}
[w_{L}]=2; W = a^{2}bcdw: \\ 
<aab|cdw> = 2(ab;acdw) + (aa;bcdw)  = 0, \\ 
<aac|bdw> = 2(ac;abdw) + (aa;bcdw)  = 0, \\ 
<aad|bcw> = 2(ad;abcw) + (aa;bcdw)  = 0, \\ 
<abc|adw> = (ab;acdw) + (ac;abdw) + (bc;aadw) = 0, \\ 
<acd|abw> = (ac;abdw) + (ad;abcw) + (cd;aabw) = 0, \\ 
<abd|acw> = (ab;acdw) + (ad;abcw) + (bd;aacw) = 0, \\ 
<bcd|aaw> = (bc;aadw) + (bd;aacw) + (cd;aabw)= 0. \\
\end{eqnarray}
These 7 equations in 7 unknowns have the solution that all unknowns 
are zero.

\begin{eqnarray}
[w_{L}]=2; W = abcdew:\label{a5} \\ 
<abc|dew> = (ab;cdew) + (bc;adew) + (ac;bdew) = 0, \\ 
nine\;\; more\;\;equations\;\; by \;\; permutations.
\end{eqnarray}
The solution of these ten simultaneous equations in ten unknows 
is that all ten are zero. We had to introduce the labels $de$ 
explicitly to get this result; which means that the tensors involved 
here are at least of third rank and also $n \ge 5$.

\section{First steps toward a general proof}

Let's start with the following  set of cases, where the explicit part 
of $W$ contains at most two distinct labels.
\begin{eqnarray}
[w_{L}] = k; \;\;\; [w_{R}] \ge k+1; \;\;\;W = a^{2k+1-m}b^{m}w; \;\;\; 0 \le m \le k \\
<a^{k+1-r}b^{r}|a^{k-m+r}b^{m-r}w> = 0 = \;\;\;\;\; \\
(k+1-r)(a^{k-r}b^{r};a^{k+1-m+r}b^{m-r}w) + 
r(a^{k+1-r}b^{r-1};a^{k-m+r}b^{m+1-r}w),
\end{eqnarray}
for $0 \le r\le m$. For each set of values for $k$ and $m$, this is a 
series of equations, 
\begin{equation}
(k+1-r)Q(r) + r Q(r-1) = 0,
\end{equation}
which leads to $Q(r) = 0$ for all allowed values of r.

Next, let's consider this set of cases: 
\begin{equation}
[w_{L}] = k; \;\;\; 	W = a^{k+1}\;x_{k}\;w, \label{d0}
\end{equation}
where $x_{k}$ is some given word of length $k$. I also introduce the notation
$x_{k,r,\alpha}$ to stand for the word 
that is made from some 
subset of $r$ letters in the word $x_{k}$; there are many such 
subsets and so the label $\alpha$ is meant to distinguish them from 
one another.Then we see the series of 
equations,
\begin{eqnarray}
<a^{k+1}|x_{k}w> = 0 = (k+1)(a^{k};a\;x_{k}w), \label{d1}\\
<a^{k}\;x_{k,1,\alpha}|a(x_{k}/x_{k,1,\alpha})w> =0 = \label{d2}\\
k(a^{k-1}\;x_{k,1,\alpha};a^{2}\;(x_{k}/x_{k,1,\alpha})w) +
(a^{k};a\;x_{k}w) \label{d3}.
\end{eqnarray}
 The quotient $x_{k}/x_{k,r,\alpha}$ stands for that word 
which results when those $r$ letters are removed from $x_{k}$.  
Eq. (\ref{d1}) involves $r=0$; and Eqs. (\ref{d2}, \ref{d3}) 
involve $r=1$ as well as $r=0$.

When we use the result of Eq. (\ref{d1}) in Eq. (\ref{d3}) we see 
that all those integral moments formed with $r=0$ and $r=1$ vanish.  
We then go on to look at $r=2$ and find that this is an inductive 
series of equations. We conclude that all integral moments built from 
the ansatz (\ref{d0}) vanish.

How does one go on to extend this proof?  If we look at some of the 
earlier examples, for example $[w_{L}]=1, \;\;W = abcw$, we see that this nice
inductive situation does not apply in all cases. 

Here is one more set of cases that we can solve analytically.
\begin{eqnarray}
[w_{L}]=k; \;\;\; W = a^{k}b^{k}cw \label{e1}\\
<a^{k-r}b^{r+1}|a^{r}b^{k-r-1}cw> = (k-r)P_{r+1}+(r+1)P_{r} = 
0,\label{e2} \\
P_{r} \equiv (a^{k-r}b^{r};a^{r}b^{k-r}cw); \;\;\;\;\;  \label{e3}\\
<a^{k-r}b^{r}c|a^{r}b^{k-r}w> = (k-r)Q_{r}+rQ_{r-1}+P_{r} = 0,\label{e4} \\
Q_{r} \equiv (a^{k-r-1}b^{r}c;a^{r+1}b^{k-r}w). \;\;\;\;\; \label{e5}
\end{eqnarray}
We can solve the Eqs. (\ref{e2}) to yield,
\begin{equation}
P_{r} = (-1)^{r}\;\frac{r!\;(k-r)!}{k!}\;P_{0}, \;\;\;\;\; r=0, k ;
\end{equation}
and also Eqs. (\ref{e4}) yield,
\begin{equation}
Q_{r} = (-1)^{r}\;\frac{r!\;(k-r-1)!}{(k-1)!}\;(-rP_{0}/k+Q_{0}), 
\;\;\;\;\; r=0, k-1 .
\end{equation}

Now, if we look at the two extreme cases for Eqs. (\ref{e4}), namely 
$r=0$ and $r=k$, we find
\begin{eqnarray}
kQ_{0} +P_{0} = 0, \\
kQ_{k-1} + P_{k} = (-1)^{k}\;(kP_{0}-kQ_{0}) = 0.
\end{eqnarray}
The solution of this is $P_{0}=Q_{0} = 0$, which makes all of the 
solutions equal to zero.

This suggests how we might solve the general problem involving at 
most three distinct labels; but it gets rather tedious.

\section{ Another special case}

Consider now the special case of W, of length (2k+1), 
consisting of all different letters. We saw examples of this in 
(\ref{c2}) and (\ref{a5}).

Let $x, y, z$ represent any of the (k+1) length words 
contained in W; we want to use these words to label the rows and 
columns of the simultaneous linear equations we are studying.
There are N of them, where N = (2k+1)!/k!(k+1)!.

For each chosen word x, there is its complement, $\bar{x} = W/x$
, a word of length k. There are also N of them.

The basic equations (\ref{a3}) are,
\begin{equation}
<x|\bar{x}> = 0 = \sum_{a}  (x/a;y=a\bar{x}) ,
\end{equation}
where a is any one of the letters 
 contained in x and y is formed by adding this letter a to $\bar{x}$. 
We can write this as the NxN system of simultaneous linear equations with 
the matrix $A_{x,y}$ whose entries are all +1 or zero.

We can see that this matrix A is symmetric. Consider the words 
$y$, 
which label the columns of $A_{x,y}$. We saw how y is  derived 
from $\bar{x} = W/x$ for each nonzero element in the row of A labeled by x. 
Consider now the row labeled by one of those y words: $A_{y,z}$.
We have the nonzero elements given by  
 $z = b \bar{y}$ for some letter b contained in y. There will be one 
 case, $b=a$,  that will yield $z= x$, exactly the word that 
y was derived from.  
So we have shown that $A_{x,y}= A_{y,x}$.

We also see that A has only zeroes on the diagonal.  So we have Tr(A) = 0.

If we look at the matrix $A^{2}$ , we see that on its diagonal will be 
the number (k+1), which is just how many $1$Õs there are in each row 
(and each column) of A.  So we conclude that $Tr(A^{2}) = N(k+1)$.

We are interested in exploring the eigenvalues, $E_{i}$,  of the matrix A. 
There are N of them and they are real numbers. 

It is easy to find one eigenvector of A. It has all entries +1 and its 
eigenvalue is (k+1).

We can also calculate (with a computer) the determinant of A, and this 
is equal to the product of all its eigenvalues.

Again, using the computer, we can search out the eigenvalues by calculating 
det(A-EI) and seeing where (and how) it goes to zero as a function of 
E. 
In the results shown in the table below, the superscript m in $(E)^{m}$ 
indicates the multiplicity of 
any eigenvalue, shown by the behavior $(E-E_{i})^{m}$ of the 
calculated determinant 
in the neighborhood of a zero.

\vskip 0.2in 
Computed properties of the matrices A

\begin{tabular}{|l|l|l|l|} \hline
	k & N & Det & Eigenvalues $\sim (E)^{m}$\\ \hline
	1 & 3 & 2 & $(+2)^{1}, (-1)^{2}$ \\ \hline
	2 & 10 & 48 &$ (+3)^{1}, (-2)^{4}, (+1)^{5}$ \\ \hline
	3 & 35 & 47,775,744 & $(+4)^{1}, (-3)^{6}, (+2)^{14}, (-1)^{14}$ \\ \hline
	4 & 126 & $10^{32.8}$ & $(+5)^{1}, (-4)^{8}, (+3)^{27}, (-2)^{48}, (+1)^{42} 
$ \\ \hline
	5 & 462 & $10^{136.4}$ & $(+6)^{1}, (-5)^{10},(+4)^{44},  (-3)^{110}, 
	(+2)^{165}, (-1)^{132}, $ \\ \hline 
    6 & 1716 & $ 10^{557.7}$ & (+7) \ldots \\ \hline 
	7 & 6435 & $10^{2259.5}$ & (+8) \ldots \\\hline 
\end{tabular}

\vskip 0.2in 

It is surprising how these results look. There are few eigenvalues; 
they are all whole numbers; and they form a neat pattern as we go 
up in k. Even the multiplicities, shown as exponents on the 
eigenvalues, may be represented by simple formulas: in 
the second column we see $m = 2k$ and in the third column 
$m=(2k+1)(k-1)$.

We shall look for more sum rules.  From any  trace formula 
we have a sum rule for the eigenvalues.
\begin{equation}
Trace (A^{r}) = \sum_{i=1,N} (E_{i})^{r} \equiv \Sigma_{r}.
\end{equation}

Let's return to the matrix $A^{2}$ and write
\begin{eqnarray}
(A^{2})_{x,z} = \sum_{y}A_{x,y}A_{y,z}, \\
y = a \bar{x} ,\; \forall\; a \in x; \;\;\;\;\; z = b \bar{y}, \; \forall\;b 
\in y.
\end{eqnarray}
There are two distinct cases. One is where $b=a$ and this is just the 
diagonal part of $A^{2}$ as earlier noted. The other is where $b \in 
\bar{x}$. This leads us to the following construction.
\begin{eqnarray}
A^{2} = (k+1)\;I + \Delta_{2} , \\
(\Delta_{2})x,z = \delta_{z, bx/a}, \;\;\; a \in x, \;\; b \in 
\bar{x},\label{g1}
\end{eqnarray}
and $I$ is the unit matrix. Reading this, it says that $\Delta_{2}$ 
connects to a new word $z$ that has one letter removed from the original 
word $x$ and replaced by a letter from the complement  $\bar{x}$.

We shall use this formula (\ref{g1}) to calculate some higher power traces.
First, however, we will need the formula,
\begin{equation}
(\Delta_{2})^{2} = C_{0}\; I + C_{2}\;\Delta_{2} + C_{4}\;\Delta_{4}.
\end{equation}

From (\ref{g1}), we count $C_{0} = k(k+1)$; and with some 
care we count $C_{2} = (2k-1)$.  $\Delta_{4}$ is a matrix that connects 
from $x$ to a word with two letters removed and replaced with two 
letters from $\bar{x}$. We count $C_{4}=4$ because there are 4 such paths; 
and the number of such paired sets is $[(k+1)k/2][k(k-1)/2]$. We also 
note that there is zero Trace for $\Delta_{2}$, $\Delta_{4}$ and also 
$\Delta_{2}\;\Delta_{4}$.

With this, we now calculate,
\begin{equation}
A^{4} = (k+1)(2k+1)\;I + (4k+1)\;\Delta_{2} + 4\;\Delta_{4};
\end{equation}
and this leads to,
\begin{eqnarray}
\Sigma_{4} =Trace(A^{4}) = N (k+1)(2k+1), \\
\Sigma_{6} =Trace(A^{6}) = N [(k+1)^{2}(2k+1) + (4k+1)k(k+1)] \\
\Sigma_{8} =Trace(A^{8}) = N [ (k+1)^{2}(2k+1)^{2}+(4k+1)^{2}k(k+1) \\
+4(k+1)k^{2}(k-1)].
\end{eqnarray}

We have verified that all the eigenvalues given in the table above do 
satisfy these summation formulas.

We can guess that the sum rules for the trace of the odd powers 
of A will be zero; but this is verified only within the limitations 
that $r < (2k+1)$.

All we really wanted here was to see that there were no eigenvalues equal 
to zero; however, what we have uncovered is quite suggestive of a 
larger mathematical reservoir hiding behind these elementary  
investigations.

\section{Discussion}

Well, this looks like there should be a general theorem and it might 
 involve something about irreducible representations of the 
permutation group.  But I don't see how to prove it.

As an example of the boundaries of this conjecture, suppose we review the 
above calculations for $[w_{L}]=2$ but limit ourselves to tensors of 
rank two. Then we find, for instance at $W=a^{2}b^{2}$,
\begin{equation}
(aa;bb) + 2(ab;ab) = 0,
\end{equation}
which tells us a relation between two integrals; but neither of them 
must be zero.

Another example: suppose the tensor is not symmetric in its indices. 
Consider $[w_{L}]=1, W=abc$ and say that the second rank tensor is 
anti-symmetric in its indices. Then one finds three equations, which 
have the solution,
\begin{equation}
(a;bc) = (b;ca) = (c;ab),
\end{equation}
but they need not vanish.

\end{document}